%% file: main.tex
%
%

	\documentclass[twocolumn,twoside]{IEEEtran}	

	\usepackage{theorem}
	\usepackage{xspace}
	\usepackage{bbold}
	\usepackage{amssymb}
	\usepackage{citesort}


	%
	\usepackage[final]{epsfig}	

	\input macro.tex

	\setlength{\voffset}{-2.54cm}
	\setlength{\hoffset}{-2.54cm}

	\setlength{\oddsidemargin}{1.5cm}
	\setlength{\evensidemargin}{1.5cm}
	\setlength{\textheight}{24cm}
	\setlength{\textwidth}{18cm}
	\setlength{\topmargin}{0.75cm}
	\setlength{\headsep}{0.75cm}
	\setlength{\headheight}{0cm}



	\def\WFou{W_{\sss NP}\XS}
	\def\FFou{F_{\sss P}\XS}
	\def\WcFou{\Wc_{\sss N}\XS}
	\def\ML{^{\sss \mathrm{ML}}}
	\def\LU{^{\sss \mathrm{L_1}}}
	\def\LD{^{\sss \mathrm{L_2}}}
	\def\SIS{^{\sss \mathrm{SIS}}}
	\def\ISD{^{\sss \mathrm{ISD}}}
	\def\RLS{_{\sss \mathrm{RLS}}}

	\def\sss{\scriptscriptstyle}
	\def\dspsty{\displaystyle}

	\newtheorem{proposition}{Proposition}
	\newtheorem{example}{Example}
	\newtheorem{remark}{Remark}

	\hyphenation{over-view} 
	\hyphenation{So-bo-lev} 

	\setlength{\itemindent}{0.5cm}
	
\bdoc

	\input text.tex
	\appendix
	\input annex.tex


\input mainTheBBL	

\edoc

%% file: text.tex
\title{Bayesian interpretation of periodograms}

\author{\JFG and \JI \thanks{\JFG and \JI are with the \ALSSun (emaux: giova@lss.supelec.fr, idier@lss.supelec.fr).}}

\markboth
{IEEE Transaction on Signal Processing}{Giovannelli and Idier: Bayesian interpretation of periodograms}

\maketitle


\begin{abstract}
The usual nonparametric approach to spectral analysis is revisited within the regularization framework. Both usual and windowed periodograms are obtained as the squared modulus of the minimizer of regularized least squares criteria. Then, particular attention is paid to their interpretation within the Bayesian statistical framework. Finally, the question of unsupervised hyperparameter and window selection is addressed. It is shown that maximum likelihood solution is both formally achievable and practically useful. 
\end{abstract}


²\section*{Notations}

\begin{tabular}{ll}
FT					& Fourier Transform															  \\
IFT				& Inverse Fourier Transform												  \\
CF					& Continuous Frequency														  \\
DF					& Discrete Frequency															  \\
UP					& Usual Periodogram															  \\
WP					& Windowed Periodogram														  \\
$L^2$				& $L_{\eC}^2([0,1])$															  \\
$H^Q$				& $H_{\eC}^Q([0,1])$															  \\
$\ell^2$			& $\ell_{\eC}^2(\eZ)$														  \\
$\Fc$				& Discrete time FT ($\ell^2 \rightarrow L^2$)						  \\
$\WcFou$			& Truncated IFT ($L^2 \rightarrow \eC^N$) 							  \\
$\WcFou^\dag$	& Adjoint operator of $\WcFou$											  \\
$\FFou$			& Square Fourier matrix ($\eC^P \rightarrow \eC^P$)				  \\
$\WFou$			& Truncated IFT matrix ($\eC^P \rightarrow \eC^N, N\leqslant P$) \\
$\WFou^\dag$	& Hermitian matrix of $\WFou$												  \\
$\eN_N$			& $\{0,1,\dots,N-1\} $														  
\end{tabular}

\section{Introduction}\label{Intro}

\PARstart{S}{pectral analysis} is a fundamental problem in signal processing. Historical papers such as~\cite{Robinson82}, tutorials such as~\cite{Kay81} and books such as~\cite{Marple87,Kay88} are evidences of the basic role of spectral analysis, whether parametric or not. 

Nonparametric approach has recently prompted renewed interest~\cite{Sacchi98} (see also~\cite{Sacchi96}) within the regularization framework and the present contribution brings a new look at these methods. It provides statistical principles rather than empirical ones in order to derive periodogram estimators. From this standpoint, the major contribution of the paper is twofold. Firstly, it proposes new coherent interpretations of existing periodograms and modern justification for windowing techniques. Secondly, it introduces a maximum likelihood method for automatic selection of the window shape. 

Moreover, \cite{Sacchi98} suffers from a twofold limitation. On the one hand, the proposed model relies on discrete frequency whereas the frequency is a continuous variable. On the other hand, restriction to separable regularization functions does not allow spectral smoothness to be accounted for. The present contribution overcomes such limitations. 

It takes advantage of a natural model in spectral analysis of complex discrete-time series: the sum of side by side pure frequencies. Two cases are investigated: 
\ben

\item the continuous frequency (CF) case which relies on an infinite number of pure frequencies $\nu\in[0,1[$ with amplitudes $a(\nu)$, $a\in L^2$

\item the discrete frequency (DF) one which relies on a finite number, say $P$ (usually large), of equally spa\-ced pu\-re fre\-quencies $\nu_p=p/P$, with amplitudes $a_p$. Let us note $\ab=\left[a_0,\dots, a_{P-1} \right]\in\eC^P$ and $\nub=\left[\nu_0,\dots, \nu_{P-1} \right]\in[0,1[^P$. 

\een

For $N$ complex observed samples $\yb=\left[y_0,\dots, y_{N-1} \right]\in\eC^N$, such models read
\beq
\begin{array}{ccll} 
\mbox{CF:}	&y_n&=& \dspsty\int_0^1 a(\nu) e^{2i\pi\nu n} \dd\nu +b_n \,,	 \cr
\mbox{DF:}	&y_n&=& P^{-1/2}\dspsty\sum_{p=0}^{P-1} a_{p} ~ e^{2i\pi p n /P}	+b_n \,,
\end{array}
\eeq
where $\bb=\left[b_0,\dots, b_{N-1} \right]\in\eC^N$ accounts for model and observation uncertainties. Let us introduce $\WcFou$ and $\WFou$:
\beq
\begin{array}{lllll} 
\mbox{CF:}	& \WcFou:	&L^2		& \longrightarrow & \eC^N \,,\cr
\mbox{DF:}	& \WFou:		&\eC^P	& \longrightarrow & \eC^N \,,
\end{array}
\eeq
the CF and DF truncated IFT so that 
\beq \label{ModelObserv}
\begin{array}{ccll} 
\mbox{CF:}	&\yb	&=& \WcFou a 	+\bb \,,	\cr
\mbox{DF:}	&\yb	&=& \WFou \ab	+\bb \,.
\end{array}
\eeq

The current problem consists in estimating the amplitudes $a$ and/or $\ab$. Thanks to the linearity of these models \wrt the amplitudes, the problem clearly falls in the class of linear estimation problems~\cite{Sorenson80,Tikhonov77,Nashed74}. But, in practice, estimation relies on a finite, maybe small, number of data $N$. As a consequence, in the CF case, a continuous frequency function $a$ lying in $L^2$ must be selected from only $N$ data. Such a problem is known to be ill-posed in the sense of Hadamard~\cite{Tikhonov77}. In the same way, under the DF formulation, since the amplitudes outnumber the available data, the problem is underdeterminate.

This kind of problem is nowadays well identified~\cite{Tikhonov77,Demoment89} and can be fruitfully tackled by means of the regularization approach. This approach rests on a compromise between fidelity to the data and fidelity to some prior information about the solution. As mentioned above, such an idea has already been introduced in several papers~\cite{Sacchi98} but also in~\cite{Kitagawa85,Wahba80,Bretthorst88,Dublanchet97}. In the autoregressive spectral estimation problem, \cite{Kitagawa85} proposes to account for spectral smoothness as a function of autoregressive coefficients. Other\-wi\-se, high resolution spectral estimation has been addressed within the regularization framework, founded on the Poisson-Gaussian model~\cite{Dublanchet97}. The present paper deepens Gaussian models and is organized as follows. 

Section~\ref{Usual} focuses on the interpretation of usual periodograms (UP) and Section~\ref{Windowed} deals with the interpretation of windowed periodograms (WP) both using penalized approaches with quadratic regularization. Results are exposed in four Propositions and the corresponding Proofs are given in Appendix~\ref{AnnexProof}. A Bayesian interpretation is presented in Section~\ref{Bayes} while the problem of parameter estimation and window selection are addressed in Section~\ref{Hyperparam}. Finally, conclusions and perspectives for future works are presented in Section~\ref{Conclu}. 
\section{Usual periodogram}\label{Usual}

\subsection{Continuous frequency}\label{SectUsualPenalCont}

The problem at stake consists in estimating $a \in L^2$ given data $\yb$ such that~(\ref{ModelObserv}). A first possible approach is founded on the Least Squares (LS) criterion
\beqx
(\yb- \WcFou a)^\dag (\yb- \WcFou a) 
= \sum_{n=0}^{N-1} \left| y_n - \int_0^1 a(\nu) e^{2i\pi\nu n} \dd\nu \right|^2 \,,
\eeqx
but, since $\WcFou$ is one-to-many and not many-to-one, there exists an infinity of solutions in $L^2$. Here, the pre\-fer\-red solution for raising the indetermination relies on Regularized Least Squares (RLS). The simplest RLS criterion is founded on quadratic ``separable regularization'':
\beq \label{CritUsualPenalCont}
\Qc_{\rm u}(a) 	= (\yb- \WcFou a)^\dag(\yb- \WcFou a)  + \lambda \int_0^1 |a(\nu)|^2 \dd\nu \,,
\eeq
where ``u'' stands for usual. The regularization parameter $\lambda\geqslant 0$ balances the trade-off between confidence in the data and confidence in the penalization term. For any $\lambda>0$, the Proposition below gives the minimizer $\est{a}^\lambda$ of~(\ref{CritUsualPenalCont}).

\medskip\begin{proposition}---\label{PropUsualPenalCont}
\textbf{(CF/UP)}.
For any $\lambda>0$, the unique minimizer of~(\ref{CritUsualPenalCont}) reads
\beq \label{EstUsualPenalCont}
\est{a}^\lambda(\nu) =  (1+\lambda)\pmu \sum_{n=0}^{N-1} y_n e^{-2i\pi\nu n} \,.
\eeq
\end{proposition}

\medskip\begin{proof}
See appendix~\ref{DemPropUsualPenalCont}.
\end{proof}\medskip

\subsection{Discrete frequency}\label{SectUsualPenalDiscr}

This subsection investigates the DF counterpart of the previous result. In the DF approach, the LS criterion reads 
\beq
(\yb - \WFou \ab)^\dag(\yb - \WFou \ab) \,,
\eeq
but, since $\WFou$ is one-to-many and not many-to-one, there also exists an infinity of solutions in $\eC^P$. According to the quadratic ``separable regularization'', the corresponding RLS criterion is
\beq \label{CritUsualPenalDiscr}
Q_{\rm u}(\ab) = (\yb - \WFou \ab)^\dag(\yb - \WFou \ab) + \lambda \ab^\dag \ab \,,
\eeq
with optimum given in the next Proposition.

\medskip\begin{proposition}---\label{PropUsualPenalDiscr}
\textbf{(DF/UP)}.
For any $\lambda>0$, the unique minimizer of~(\ref{CritUsualPenalDiscr}) reads
\beq \label{EstUsualPenalDiscr}
\est{\ab}^\lambda = (1+\lambda)\pmu \FFou \tilde{\yb}_P \,,
\eeq
where $\tilde{\yb}_P$ denotes the vector $\yb$ \textit{zero-padded} up to size $P$. 
\end{proposition}

\medskip\begin{proof}
See appendix~\ref{DemPropUsualPenalDiscr}.
\end{proof}\medskip

\subsection{Usual periodogram: concluding remarks}

In the CF cases, the squared modulus of the penalized solutions $|\est{a}^\lambda(\nu)|^2$ is proportional to the usual zero-padded periodogram. Moreover, $|\est{\ab}^\lambda|^2$ is\footnote{If $\ub\in\eC^P$, $|\ub|^2$ denotes the vector of the squared moduli of the component of $\ub$.} a discretized version of $|\est{a}^\lambda(\nu)|^2$ over the frequency grid $\nub$. So, within the proposed framework, \textit{separable quadratic regularization} leads to the \textit{usual zero-padding} technique associated with the practical computation of periodograms. Moreover, when $\lambda$ tends to zero, the proportionality factor tends to one. It is noticeable that, in this case, the criteria~(\ref{CritUsualPenalCont}) and~(\ref{CritUsualPenalDiscr}) degenerate but their minimizer does not: they are the solution of the constraint problems 
\beqx
\begin{array}{llll} 
\mbox{CF:}	& \dspsty \min_{a\in L^2} \int_0^1 |a(\nu)|^2 \dd\nu	&\mathrm{s.t.}	& \yb=\WcFou a	\,, 	\cr 
\cr
\mbox{DF:}	& \dspsty \min_{\ab \in {\eC}^P} \ab^\dag \ab			&\mathrm{s.t.}	& \yb = \WFou \ab \,.
\end{array}
\eeqx
\ie solution of the noiseless problems adressed in~\cite{Sacchi96,Sacchi98}.
\section{Windowed periodogram}
\label{Windowed}

The previous section investigates the relationships between the separable regularizers and the usual (non-windowed) periodograms. The present section focuses on smoothing regularizers and windowed periodograms (see~\cite{Harris78} which analyzes dozens of windows to compute smoothed periodograms). 

\subsection{Continuous spectra}

This subsection generalizes the usual norm in $L^2$ to the Sobolev~\cite{Bertin93} regularizer: 
\beqx
\Rc_Q(a) = \int_0^1 \sum_{q=0}^{Q} \alpha_q  \left| \frac{\dd^q a}{\dd \nu^q} (\nu) \right|^2 \dd \nu\,,
\eeqx
which can be interpreted as a measure of spectral smoothness. The $\alpha_q$ are positive real coefficients and can be generalized to positive real functions~\cite{Tikhonov77}. $\Rc_Q$ is defined onto the Sobolev space~\cite{Bertin93} $H^Q \subset L^2$. Note that $H^0=L^2$ and that the usual norm invoked in subsection~\ref{SectUsualPenalCont} is the regularizer $\Rc_0$ with $\alpha_0=1$. 

\begin{remark}---
Strictly speaking, $\Rc_Q(a)$ is not a spectral smoothness measure, since it is not a function of $|a(\nu)|$ but a function of $a(\nu)$, including phase. A true spectral smoothness measure does not depend on the phase of $a(\nu)$ and does not yield a quadratic criterion. The same remark holds for the definition of spectral smoothness proposed by Kitagawa and Gersh~\cite{Kitagawa85}.
\end{remark}

Accounting for spectral smoothness by means of $\Rc_Q(a)$ yields a new penalized criterion
\beq \label{CritWinPenalCont}
\Qc_{\rm s}(a) = (\yb- \WcFou a)^\dag (\yb- \WcFou a) + \lambda \Rc_Q(a) \,,
\eeq 
where the index ``s'' stands for smoothness. 

\medskip\begin{proposition}---\label{PropWinPenalCont}
\textbf{(CF/WP)}.
With the previous notations and definitions, the minimizer of~(\ref{CritWinPenalCont}) reads
\beq \label{EstWinPenalCont}
\est{a}^\omega(\nu)= \sum_{n=0}^{N-1} \omega_n y_n e^{-2i\pi\nu n} \,,
\eeq
\ie a windowed FT. The window shape is 
\beqn 
\omega_n &=& (1+\lambda \varepsilon_n)\pmu \,,\label{FenWinPenalCont} \\
\WITH ~~\varepsilon_p &=& \sum_{q=0}^{Q} \alpha_q (2\pi p)^{2q} \mbox{~~~for~} p\in \eZ\,.\label{DefVPWinPenalCont}
\eeqn
\end{proposition}

\medskip\begin{proof}
See appendix~\ref{DemPropWinPenalCont}.
\end{proof}\medskip

\subsection{Discretized spectra}

This subsection is devoted to the generalization of criterion~(\ref{CritUsualPenalDiscr}) to non-separable penalization
\beq \label{CritWinPenalDiscr}
Q_{\rm s}(\ab) = (\yb - \WFou \ab)^\dag(\yb - \WFou \ab) + \lambda \ab^\dag \Pi_a \ab \,.
\eeq
Given that the sought spectrum is circular-periodic, the penalization term has to be designed under circularity constraint. As a consequence, $\Pi_a$ is a circular matrix, its eigenvalues, denoted $e_p, p\in\eN_P$, can be calculated as the FT of the first row of $\Pi_a$. Moreover, without loss of generality, we assume that the diagonal elements of $\Pi_a\pmu$ are equal to one and any scaling factor is integrated in the parameter $\lambda$. 

\medskip\begin{proposition}---\label{PropWinPenalDiscr}
\textbf{(DF/WP)}.
The minimizer of~(\ref{CritWinPenalDiscr}) reads
\beq \label{EstWinPenalDiscr}
\est{\ab}^w = \FFou \breve{\yb} \,,
\eeq
where the $\breve{y}_p=w_p\tilde{y}_p$ for $p\in\eN_P$ and
\beqx
w_p = (1 + \lambda e_p)\pmu \,.
\eeqx
\end{proposition}

\medskip\begin{proof}
See appendix~\ref{DemPropWinPenalDiscr}.
\end{proof}\medskip

\subsection{Windowed periodograms: concluding remarks}

Hence, in the CF case, the squared modulus of the penalized solution $\est{a}^\omega$ is the 
windowed periodogram associated with window $\omega_n$. Moreover, the DF solution $\est{\ab}^w$ is a discretized version of $\est{a}^\omega$, as soon as the $e_n$ are identified with the $\varepsilon_n$. As a conclusion \textit{quadratic smoothing regularizers} interpret \textit{windowed periodograms}. Moreover, it is noteworthy that $\est{a}^\omega(\nu)$ and $\est{\ab}^w$ only depend on $e_n$ and $\varepsilon_n$ for $n\in \eN_N$. 

\begin{remark}--- 
\textbf{Empirical power.} \label{PuissCFPMWP}
One can easily show:
\beq
\begin{array}{lccl} 
\mbox{CF:}	&\dspsty \int_0^1|a(\nu)|^2 \dd\nu	&=& \sum_{n=0}^{N-1} \omega_n^2 |y_n|^2 \,,\cr
\mbox{DF:}	&\dspsty \ab^\dag \ab					&=& \sum_{n=0}^{N-1} w_n^2 |y_n|^2 \,.
\end{array}
\eeq
Hence, the empirical power of the estimated spectra is smaller than the empirical power of the observed data and equality holds if and only if $\lambda=0$.
\end{remark}

\begin{example}--- \label{ExWinPenalCont0}
\textbf{Zero-order penalization}.
The most simple example consists in retrieving the non-windowed case of section~\ref{SectUsualPenalCont} and~\ref{SectUsualPenalDiscr}. Let us apply the previous Propositions~\ref{PropWinPenalCont} and~\ref{PropWinPenalDiscr} with regularizers
\beq
\begin{array}{lccl} 
\mbox{CF:}	&\dspsty \int_0^1|a(\nu)|^2 \dd\nu	&\ie& Q=0 ~~\AND~ \alpha_0=1 \,, \cr
\mbox{DF:}	&\dspsty \ab^\dag \ab					&\ie& \Pi_a = I_P \,.
\end{array}
\eeq
Then, we have $\varepsilon_n=e_n=1$, the criteria~(\ref{CritWinPenalCont}) and~(\ref{CritWinPenalDiscr}) res\-pec\-ti\-ve\-ly become~(\ref{CritUsualPenalCont}) and~(\ref{CritUsualPenalDiscr}) and the solutions~(\ref{EstWinPenalCont}) and~(\ref{EstWinPenalDiscr}) respectively become~(\ref{EstUsualPenalCont}) and~(\ref{EstUsualPenalDiscr}): as expected, the non-windowed solutions are retrieved. A more interesting example is the one given below. 
\end{example}

\begin{example}--- \label{ExDiscrPenalDer}\label{ExWinPenalCont1}
\textbf{First-order penalization.}
Let the penalization term be 
\beq
\begin{array}{ll} 
\mbox{CF:}	&\dspsty \int_0^1|a'(\nu)|^2 \dd\nu	\,,\cr
\mbox{DF:}	&\dspsty \frac{1}{2}P^2\sum_{k=0}^P |a_k-a_{k-1}|^2 \,.
\end{array}
\eeq
with $a_P=a_0$ for notational convenience of the circularity assumption. 
Application of Propositions~\ref{PropWinPenalCont} and~\ref{PropWinPenalDiscr} respectively yields $\varepsilon_n=4\pi^2 n^2$ (CF case) and $e_n=(1-\cos 2\pi n/P)$ (DF case). The corresponding windows read
\beq
\begin{array}{ccll} 
\mbox{CF:}	&\omega_n	&=& (1+4\pi^2n^2\lambda)\pmu \,, \cr
\mbox{DF:}	&w_n	&=& (1 +\lambda - \lambda \cos 2\pi n/P)\pmu \,.
\end{array}
\eeq
In the following, we refer to them as the Cauchy and the inverse cosine windows. Moreover, for a finer discretization of the spectral domain, $\lim_{P\rightarrow\infty}e_n=\varepsilon_n$ and one can retrieve the Cauchy window as the limit of the inverse cosine window.
%
\begin{figure}[htbp]
\centerline{\psfig{file=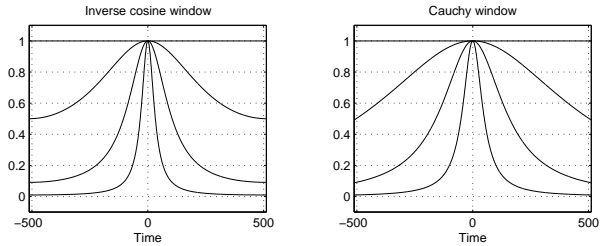,width=8cm}}
\caption[Inverse cosine and Cauchy windows]{Inverse cosine window (lhs) and Cauchy window (rhs) as a function of $\lambda$. In both cases, $\lambda=0$ yields a constant shape. Furthermore, for any $\lambda$, $\omega_0=w_0=1$. Otherwise, as $\lambda$ increases the window shape decreases faster to zero and the corresponding spectrum is smoothed.}
\label{FigCauchCosWin}
\end{figure}
%
\end{example}

\section{Bayesian interpretation}
\label{Bayes}

This section is devoted to Bayesian interpretations of the penalized solutions presented in Propositions~\ref{PropUsualPenalCont}, \ref{PropUsualPenalDiscr}, \ref{PropWinPenalCont} and~\ref{PropWinPenalDiscr}. Moreover, since usual non-windowed forms are particular cases of windowed forms, we focus on the latter. 

Since the considered criteria are quadratic, their Bayesian interpretations rely on Gaussian laws. Therefore, the Bayesian interpretations only require the characterization of means and correlation structures for the stochastic models at work. 

\subsection{Discrete frequency approach}

In the DF case, \ie in the finite dimension vector space, the Bayesian interpretation of the criteria~(\ref{CritUsualPenalDiscr}) and~(\ref{CritWinPenalDiscr}) as a \post Co-Log-Likelihood is a classical result~\cite{Demoment89}. Within this probabilistic framework, the likelihood of the parameters $\ab$ attached to the data $\yb$ is
\beqx
f(\yb \vert \ab) = (\pi r_b)^{-N} \exp{ \frac{-1}{r_b} (\yb - \WFou \ab)^\dag(\yb - \WFou \ab) } \,.
\eeqx
From a statistical viewpoint, it essentially results from the linearity of the model~(\ref{ModelObserv}) and from the hypothesis of a zero-mean, circular (in the statistical sense), stationary, white and Gaussian noise vector $\bb$, with variance $r_b$.

Moreover, in order to interpret the regularization term of~(\ref{CritWinPenalDiscr}), a zero-mean, circular, correlated Gaussian prior with covariance $R_a=r_a\Pi^{-1}_a$ is introduced\footnote{Rigorously speaking, this is possible only if $\Pi_a$ is invertible.}. Matrix $\Pi^{-1}_a$ is the normalized covariance structure, \ie all its diagonal elements are equal to~1, while $r_a$ stands for the prior power. So, the prior density reads 
\beqx
f(\ab) = ( \pi r_a)^{-N} \det{\Pi_a} \exp{ \frac{-1}{r_a} \ab^\dag \Pi_a \ab } \,.
\eeqx
%

\begin{figure}[htbp]
\centerline{\psfig{file=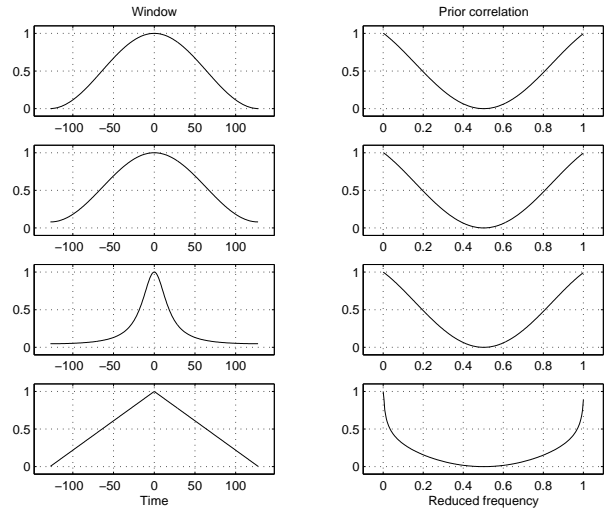,width=8cm}}
 \caption[Usual windows and the corresponding correlations]{Usual windows and the corresponding correlations. The lhs column shows the time window and the rhs column shows the associated correlations. From top to bottom: the Hamming, the Hanning, the inverse cosine and the triangular.}
 \label{FigFenCov}
\end{figure}

The Bayes rule ensures the fusion of the likelihood and the prior into the \post density
\beqx
f(\ab \vert \yb) \propto \exp{ \frac{-1}{r_b}Q_{\rm s}(\ab) } \,,
\eeqx
where $Q_{\rm s}$ is given by Eq.~(\ref{CritWinPenalDiscr}). The regularization parameter $\lambda$  is clearly $\lambda=r_b/r_a$. 

Thus, we have a Bayesian interpretation of the criterion~(\ref{CritWinPenalDiscr}) related to windowed periodograms. Interpretation of the criterion~(\ref{CritUsualPenalDiscr}) related to usual ones results from a white prior: $\Pi_a=I_P$. Finally, interpretations of the RLS solutions~(\ref{EstUsualPenalDiscr}) and~(\ref{EstWinPenalDiscr}) themselves, result from the choice of the Maximum \textit{A Posteriori} (MAP) as a punctual estimate. Moreover, thanks to the Gaussian character of \post law, other basic Bayesian estimators such as \Post Mean (PM) and Marginal MAP (MMAP), are equal to the MAP solution itself.

\subsection{Continuous frequency case}

\subsubsection{General theory}

In the CF case, the Bayesian interpretation is more subtle since it relies on continuous index stochastic processes. Indeed, no \post likelihood for the parameter $a$ is available. So, there is no direct \post interpretation of the criteria~(\ref{CritUsualPenalCont}) and~(\ref{CritWinPenalCont}), nor MAP interpretation of the estimates~(\ref{EstUsualPenalCont}) and~(\ref{EstWinPenalCont}). Roughly speaking, the \post law vanishes everywhere. Nevertheless, there is a proper Bayesian interpretation of the estimates~(\ref{EstUsualPenalCont}) and~(\ref{EstWinPenalCont}) as PM or MMAP as shown below. 


Let us introduce a zero-mean, circular (in the statistical sense) and Gaussian prior law~\cite{Cramer67} for $a$. This law is fully characterized by its correlation structure $\gamma_a(\nu),\nu\in[-1,1]$, which is entirely described by its values for $\nu\in [0,1]$ thanks to Hermitian symmetry. Furthermore, the usual circular-periodicity assumption for $a(\nu)$ results in another symmetry property: $\gamma_a(1/2+\nu) = \gamma_a(1/2-\nu)$ for any $\nu\in[0,1/2]$. 

By assuming $\gamma_a\in L_2$, the latter can be expanded into a Fourier series: 
\beqx
\gamma_a(\nu) = \sum_{p\in \eZ} \rond{\gamma_a}(p) e^{-2i\pi\nu p} \,,\, \nu \in[0,1]
\eeqx
with Fourier coefficients $\rond{\gamma_a}\in \ell_2$ given by:
\beqx
\rond{\gamma_a}(p) = \int_{[0,1]} \gamma_a(\nu)e^{-2i\pi\nu p} \,,\, p \in \eZ \,.
\eeqx
Let us note $c_a(\nu)=\gamma_a(\nu)/r_a$ the normalized correlation and $\rond{c_a}\in \ell_2$ the corresponding Fourier sequence. 

\medskip\begin{proposition}---\label{PropBayesCont}
With the previous notations and prior choice, the \post mean of $a(\nu)$ is:
\beqn 
\Esp{a(\nu) | \yb}
			&=& \est{a}^\omega(\nu) ~=~ \sum_{n=0}^{N-1} \omega_n y_n e^{-2i\pi\nu n} \,,\label{PMBayesWinPenalCont} \\
\WITH~ \omega_n	
			&=& \left[ 1+\lambda \rond{c_a}(n)\pmu \right]\pmu \,. \label{FenBayesWinPenalCont}
\eeqn
\end{proposition}

\medskip\begin{proof}
See appendix~\ref{DemPropBayesCont}.
\end{proof}\medskip

Comparison of~(\ref{PMBayesWinPenalCont})-(\ref{FenBayesWinPenalCont}) and~(\ref{EstWinPenalCont})-(\ref{FenWinPenalCont}) immediately gives the Bayesian interpretation of windowed FT as PM\footnote{Since $a(\nu)|\yb$ is a scalar Gaussian random variable $\Esp{a(\nu) | \yb}$ is also the MMAP.}: $\rond{c_a}(n)=\varepsilon_n\pmu$, \ie identification of the Fourier coefficients of the prior correlation $c_a(\nu)$ and the FT of the discrete correlation $\Pi_a$.

\addtocounter{example}{1}
\subsubsection{Example \arabic{example}} \label{ExWinPenalContBayes}

The present subsection is devoted to a precise Bayesian interpretation of deterministic examples~\ref{ExWinPenalCont0} and~\ref{ExWinPenalCont1}. As we will see, there is a new obstacle in the Bayesian interpretation of these examples because the underlying correlations do not lie in $L_2$. In order to overcome this difficulty we first interpret the penalization of both zero-order and first-order derivative:
\beq \label{PenalDoux01}
\Rc_2(a) = \alpha_0 \int_0^1 \left|a(\nu)\right|^2 \dd \nu + \alpha_1 \int_0^1 \left|a'(\nu)\right|^2 (\nu)\dd \nu \,.
\eeq
The case of pure zero-order and pure first-order are obtained in sections~\ref{Limitalpha1} and~\ref{Limitalpha0} as limit processes.

As seen in Proposition~\ref{PropWinPenalCont}, Eq.~(\ref{DefVPWinPenalCont}), the associated coefficients are: $\varepsilon_p = \alpha_0 + 4 \pi^2 \alpha_1 p^2 \,,\, p\in \eZ$. According to Proposition~\ref{PropBayesCont}, the Fourier series coefficients for $\gamma_a(\nu)$ are $\rond{\gamma_a}(p)=\varepsilon_p\pmu$. It is clear that $\rond{\gamma_a} \in \ell_2$, hence $\gamma_a\in L_2$ and 
\beq \label{TheFourierSeries}
\gamma_a(\nu) = \sum_{p\in\eZ} \frac{1}{\alpha_0+4\pi^2\alpha_1 p^2}~e^{-2i\pi\nu p} \,,\, \nu \in[0,1] \,.
\eeq
It is shown in Appendix~\ref{AnnexeFouSerie} that, with $\alpha=\sqrt{\alpha_0/\alpha_1}$ and $\alpha'=\sqrt{\alpha_0\alpha_1}$, $\gamma_a(\nu)$ reads
\beq \label{Cov01}
\gamma_a(\nu) = \frac{\cosh{\alpha(|\nu|-1/2)}}{2\alpha' \sinh{\alpha/2}} \,,\, \nu \in[-1,1] \,,
\eeq
and several analytic properties are straightforwardly deduced. In particular, $\gamma_a$ has a continuous derivative over $[-1,1]-\{0\}$ and the slopes at $\nu=0^-$ and $\nu=0^+$ are respectively $1/\alpha_1$ and $-1/\alpha_1$. $\gamma_a$ is minimum at $\nu=1/2$ and maximum at $\nu=-1$, $\nu=0$ and $\nu=1$. Moreover its integral from 0 to 1 remains constant and equals $1/\alpha_0$. 

\paragraph{Markov property} The present paragraph addres\-ses the Markov property of the underlying prior process $a(\nu)$~\cite{Bremaud99,Moura97}. This process cannot be \strsen a Mar\-kov chain since it is circular-periodic: ``future'' frequency and ``past'' frequency cannot be independent. However, we show the Markov property for the conditional process $\bar{a}(\nu)=[a(\nu)|a(1)]_{\nu\in[0,1[}$. It is shown in Appendix~\ref{AnnexeCondProc} that its correlation structure reads
\beqn
\gamma_{\bar{a}}(\nu,\nu')
	&=&\gamma_a(\nu-\nu')-\frac{\gamma_a(\nu)\gamma_a(\nu')}{\gamma_a(0)}				\label{CondCorrel1}\\
	&=&\frac{\sinh{\alpha \nu'} ~ \sinh{\alpha (1-\nu})}{\alpha' \sinh{\alpha}} \,,	\label{CondCorrel2}
\eeqn
for any $\nu,\nu'\in[0,1], \nu\geqslant\nu'$. According to the sufficient factorization of the correlation function proposed in~\cite[p.64]{Wong71}, it turns out that $\bar{a}(\nu)$ is a Markov chain.

\paragraph{Limit case as $\alpha_1\rightarrow0$}\label{Limitalpha1}As $\alpha_1$ tends to zero, it is easy to show that for each $\nu\in ]0,1[$, the correlation $\gamma_a(\nu)$ tends to zero \ie there is no more correlation between $a(\nu_1)$ and $a(\nu_2)$ as soon as $\nu_1\ne\nu_2$ and $(\nu_1,\nu_2)\ne(0,1)$. Moreover, $\gamma_a(0)$ and $\gamma_a(1)$ tend to infinity while the integral of $\gamma_a$ over $[0,1]$ remains $1/\alpha_0$. Roughly speaking, the limit correlation is a Dirac distribution at $\nu=0$ and $\nu=1$ with weight $1/2\alpha_0$ \ie the limit process is a circular white Gaussian noise with ``pseudo-power'' $1/\alpha_0$.

\paragraph{Limit case as $\alpha_0\rightarrow0$}\label{Limitalpha0}This case is more complex than the previous one since $\forall \nu \in [0,1]$, $\gamma_a(\nu)$ tends to infinity as $\alpha_0$ tends to zero. So, we propose a characterization of the limit process \textit{via} its increments. Let $\nu_1, \nu_2, \nu'_1, \nu'_2 \in [0,1]$, $\nu_1 < \nu_2 < \nu'_1 < \nu'_2$. Let us also note the frequency increments $\tau_{\nu}=\nu_2-\nu_1$ and $\tau'_{\nu}=\nu'_2-\nu'_1$ and the vector of the increments themselves $\ib=[a(\nu_2)-a(\nu_1), a(\nu_4)-a(\nu_3)] \in \eC^2$. This vector is clearly Gaussian and zero-mean. Furthermore, it is shown in Appendix~\ref{AnnexeIncrements} that its covariance matrix reads
\beq \label{IncreCov}
R_i = \frac{1}{2\alpha_1}
\left[\begin{matrix} 
	\tau_{\nu} (1-\tau_{\nu})	& 2\tau_{\nu}	\tau'_{\nu}		\\
	2\tau_{\nu}	\tau'_{\nu}		& \tau'_{\nu} (1-\tau'_{\nu})
\end{matrix}\right] \,.
\eeq
It turns out that the process $\tilde{a}(\nu)=a(\nu)-a(0)$ is a Brownian bridge~\cite[p.36]{Bhattacharya90}. 
\section{Hyperparameter and window selection}
\label{Hyperparam}

The problem of hyperparameter estimation within the regularization framework is a delicate one. It has been extensively studied and numerous techniques have been proposed and compared~\cite{Golub79,Titterington85,Hall87,Thompson91,Fortier92,Giovannelli96}. The Maximum Likelihood (ML) approach is often chosen associated with the Bayesian interpretation. In the following subsections, we address regularization parameter estimation and automatic window selection using ML estimation.

\subsection{Hyperparameters estimation}

In our context, the ML technique consists in integrating the amplitudes out of the problem and maximizing the resulting marginal likelihood \wrt the hyperparameters. Thanks to the linear and Gaussian assumptions, the marginal law for the data, namely the likelihood function, is also Gaussian
\beq \label{HyperVrais}
f(\yb \,;\, r_a,r_b) \propto (\det{ R_{\yb} })\pmu \exp{ - {\yb}^\dag R_{\yb}\pmu \yb} \,.
\eeq
Moreover, the covariance structure $R_\yb$ can be easily derived, as shown in the two following sections. 

\subsubsection{Discrete frequency marginal covariance}

In the present case, since all random quantities are in a finite dimensional linear space, the covariance is clearly 
\beqx
R_\yb = r_a ( \WFou \Pi^{-1}_a \WFou^\dag + \lambda I_N ) = r_a \Sigma_\yb \,.
\eeqx
Accounting for the circular structure of the matrix $\Pi_a$, we have $\Pi_a=\FFou\Lambda_\Pi\FFou^\dag$, where $\Lambda_\Pi$ is the diagonal matrix of eigenvalues: $e_p, p\in \eN_P$. Given the property~(\ref{WFouFFou}) in Appendix~\ref{AnnexTech}, $\Sigma_\yb$ is shown to be diagonal
\beq
\Sigma_\yb = \Diag{\lambda + e_n\pmu}, \, n\in \eN_N \,.
\eeq

\subsubsection{Continuous frequency marginal covariance}

In the present case, the marginal covariance matrix $R_\yb$ has already been derived in Appendix~\ref{DemPropBayesCont}, Eq.~(\ref{Rnm}). Hence, $R_{\yb}$ and $\Sigma_\yb$ are diagonal:
\beq
\Sigma_\yb = \frac{1}{r_a} R_\yb = \Diag{\lambda + \varepsilon_n\pmu}, \, n\in \eN_N \,.
\eeq

\begin{remark}--- \label{RemVraisDiscr}
In both cases, $\Sigma_\yb$ only depends on $e_n$/$\varepsilon_n$ for $n\in \eN_N$. Consequently the likelihood function and the ML parameter only depend on the $N$ first coefficients. 
\end{remark}

\subsubsection{Maximization}

The opposite of the logarithm of the likelihood, namely the Co-Log-Likelihood (CLL)
\beq \label{PerioALVHyper}
CLL(r_a,\lambda) = N \log{r_a} + \log{ \det{\Sigma_\yb} } + \frac{1}{r_a} \yb^\dag \Sigma_\yb\pmu \yb \,,
\eeq
must be minimized \wrt~$r_a$ and $\lambda$. Partial minimization is tractable \wrt~$r_a$ and yields $\est{r}_a = \yb^\dag \Sigma_\yb\pmu \yb/N$. Substitution of $\est{r}_a$ in Eq.~(\ref{PerioALVHyper}) gives:
\beq \label{PerioALVLambda}
CLL(\lambda) = \log{ \det{\Sigma_\yb} } + N \log{ \yb^\dag \Sigma_\yb\pmu \yb } \,.
\eeq
Furthermore, since $\Sigma_\yb$ is a diagonal matrix
\beqnx
CLL(\lambda) 
	&=& \sum_{n=1}^{N} \log{(\lambda+ e_n\pmu)} 
	+ N \log{ \sum_{n=1}^{N} \frac{|y_n|^2}{\lambda+ e_n\pmu} } \\
	&=& \log{\left\{ \prod_{n=1}^{N} (\lambda+ e_n\pmu) 
	 \left[ \sum_{n=1}^{N} \frac{|y_n|^2}{\lambda+ e_n\pmu} \right]^N \right\}} \,,
\eeqnx
in the DF case. Substitution of $e_n$ by $\varepsilon_n$ yields the CF case. In both cases, $CLL(\lambda)$ is the logarithm of the ratio of two degree $N-1$ polynomials of the variable $\lambda$, with a strictly positive denominator. Minimization \wrt~$\lambda$ is not explicit, but it can be numerically performed.

\subsubsection{Simulation results}

ML hyperparameter selection is illustrated for the problem of Section~\ref{ExWinPenalContBayes}. Computations have been performed on the basis of  of 512 sample signals simulated by filtering standard Gaussian noises with the filter of impulse response $h=[1,-2,3,-2,1]$. Let us note $a^\star$ as the true spectrum. 

CLL has been computed on a $(\alpha_0,\alpha_1)$-grid of $100 \times 100$  logarithmically spaced values from $10^{-10}$ to $10^{10}$. The first observation is that CLL is fairly regular and usually shows a unique minimum, located between $10^{-1}$ and $10^{1}$ for $\alpha_0$, and between $10^{-2}$ and $1$ for $\alpha_1$. However, a few ``degenerated'' cases have been observed for which $\est{\alpha}_0\ML$ or $\est{\alpha}_1\ML$ seem to be null or infinite. Let us note $(\est{\alpha}_0\ML,\est{\alpha}_1\ML)$ as the CLL minimizer\footnote{Efficient algorithms are available in order to maximize the likelihood, such as gradient based~\cite{Bertsekas95} or EM type~\cite{Shumway82}. They have not been implemented here as far as a mere feasibility study is concerned.} and $\est{a}\ML\RLS$ as the corresponding RLS periodogram.

Since $a^\star$ is known in the proposed simulation study, various spectral distances~\cite{Basseville89} can be computed, as functions of $\alpha_0$ and $\alpha_1$. $L_1$ distance, $L_2$ distance, the Itakura-Saito  divergence (ISD) as well as the Itakura-Saito symmetric distance (SIS) have been considered. Each one provides an optimal couple 
$(\est{\alpha}_0\LU,\est{\alpha}_1\LU)$,
$(\est{\alpha}_0\LD,\est{\alpha}_1\LD)$, 
$(\est{\alpha}_0\ISD,\est{\alpha}_1\ISD)$, and
$(\est{\alpha}_0\SIS,\est{\alpha}_1\SIS)$ respectively. The corresponding spectra are respectively denoted 
$\est{a}\LU\RLS$, 
$\est{a}\LD\RLS$,
$\est{a}\ISD\RLS$, and
$\est{a}\SIS\RLS$.

\begin{figure}[htbp] 
\centerline{\psfig{file=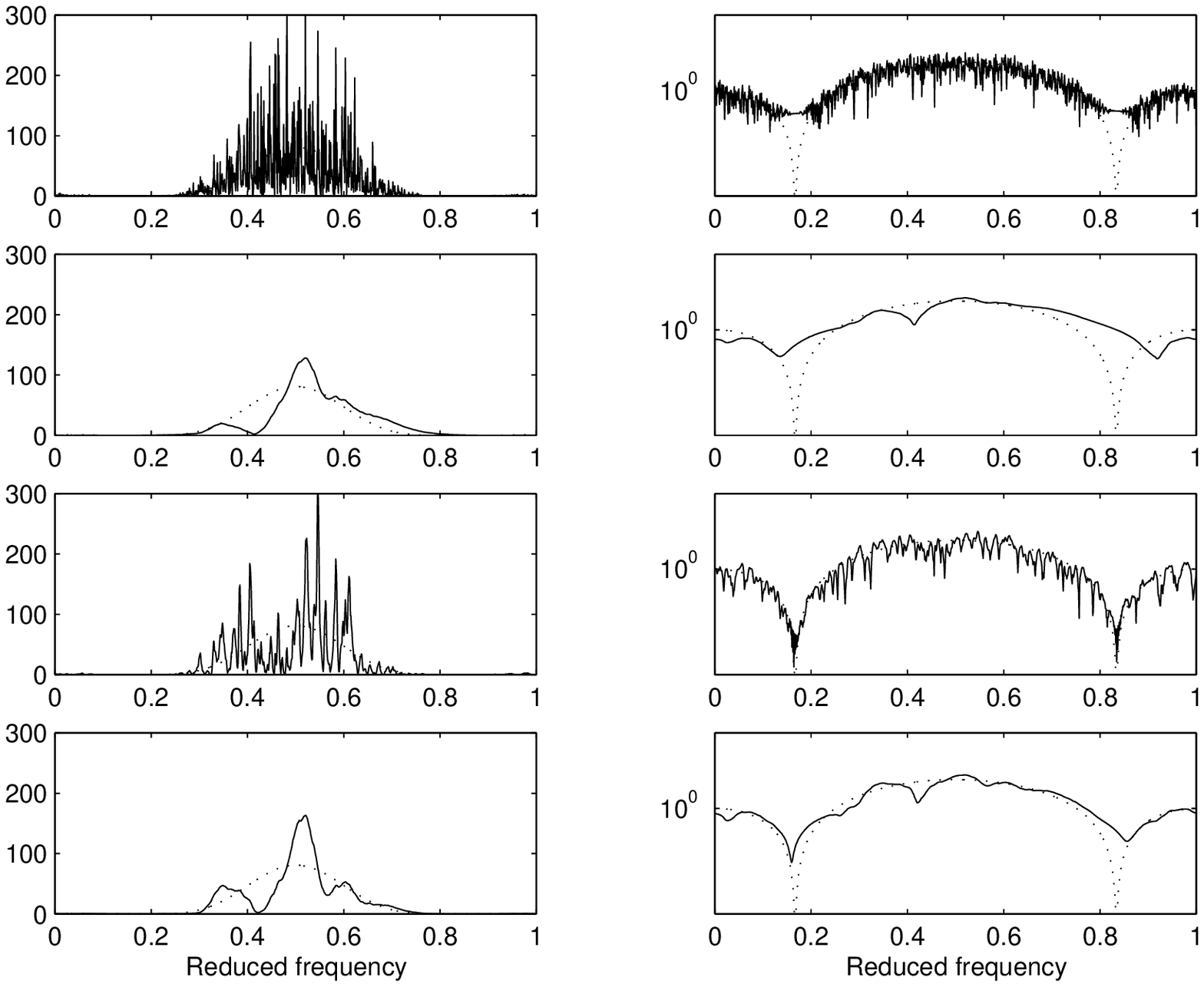,width=8cm}}
\caption[Qualitative comparison]{Qualitative comparison. True spectra (dotted lines) and estimated ones (solid lines). The lhs column gives linear plots and the rhs column gives logarithmic plots. From top to bottom\footnote{$\est{a}\LU\RLS$ and $\est{a}\SIS\RLS$ have also been computed. They are not reported here since they respectively behave akin to $L_2$ and Itakura-Saito divergence.}: usual periodograms, $\est{a}\LD\RLS$, $\est{a}\ISD\RLS$ and $\est{a}\ML\RLS$.}
\label{ResPerio}
\end{figure}

According to our experiments, as shown in Fig.~\ref{ResPerio}, $\est{a}\LD\RLS$, $\est{a}\ISD\RLS$ and the $a^\star$ can be graded by smoothness and estimation accuracy. From the smoothest to the roughest, the following gradation has always been observed: $\est{a}\LD\RLS$, $a^\star$ and $\est{a}\ISD\RLS$. Furthermore, $\est{a}\LD\RLS$ is systematically over-smoothed while $\est{a}\ISD\RLS$ is systematically under-smoothed. Moreover, the first one qualitatively approximates more precisely $a^\star$ in linear scale, whereas the second one reproduces more accurately $a^\star$ in a logarithmic scale and especially the two notches. This is due to the presence of the spectra ratio in the Itakura-Saito distance which emphasizes the small values of the spectra. 

Finally, to our experience and as shown in Fig.~\ref{ResPerio}, the maximum likelihood solution $\est{a}\ML\RLS$ establishes a relevant compromise between $\est{a}\LD\RLS$ and $\est{a}\ISD\RLS$ since it is smooth enough, while the two notches remain accurately described. 

Quantitative comparisons have been conducted between the two practicable methods (when $a^\star$ is not known): the usual periodogram and the proposed method \ie the RLS solution with automatic ML hyperparameters. The obtained results are reported in Table~\ref{TableDist}. They clearly show an improvement of about 40-50\% for all the considered distances. 
\begin{table}[htbp] 
\begin{center}\begin{tabular}{|c|c|c|c|c|} 							\hline 
				& $L_1$		& $L_2$		& $AIS$		& $SIS$			\\ \hline 
UP				& 0.766		& 1.14~~		&	751		& 	750			\\
RLS + ML		& 0.471		& 0.567		&	420 		& 	422 			\\ \hline 
Gain			& 38.5\%		& 50.3\%		&	44.1\%	& 	43.8\%		\\ 
																					\hline
\end{tabular} \end{center} 
\caption[Quantitative comparison]{Quantitative comparison. The first line refers to the usual periodogram while the second one refers to the RLS solution with ML hyperparameters. The third line gives the quantitative improvement.}
\label{TableDist}
\end{table}

\subsection{Window selection}

It has been shown that the ML technique allows the estimation of the regularization parameter. The problem of window selection is now addressed. Let us consider a set of $K$ windows \ie $K$ matrices $\Pi_a^k$ for $k \in \eN_K$. Index $k$ becomes a new hyperparameter as well as $\lambda$, and can be jointly estimated. The likelihood function~(\ref{PerioALVLambda}) is now
\beqx
CLL(\lambda, k) = 
 \log{ \det{(\Sigma_\yb^k)} } + \log{ N \yb^\dag (\Sigma_\yb^k)\pmu \yb} \,.
\eeqx
Maximization \wrt hyperparameters can be achieved in the same way as above for each value of $k\in\eN_K$. The maximum maximorum can then be easily selected. 

Numerous simulations have been performed. They are not reported here since they show similar results as the previous ones. However, it has been observed that the triangular window is the most often selected among: Cauchy, inverse cosine, Hanning, Hamming and triangle. 
\section{Conclusion}
\label{Conclu}

In this paper, the usual nonparametric approach to spectral analysis has been revisited within the regularization framework. We have shown that usual and windowed periodograms could be obtained \via the minimizer of regularized least squares criteria. In turn, penalized quadratic criteria are interpreted within the Bayesian framework, so that periodograms are interpreted \via Bayesian estimators. The corresponding prior is a zero-mean Gaussian process, fully specified by its correlation function. Particular attention is paid to the connection between correlation structure and window shape. As regards \emph{quadratic} regularization, the present study significantly deepens a recent contribution by Sacchi \etal~\cite{Sacchi98}, given that the latter addresses neither windowed periodograms, nor the continuous frequencial setting. Extension to the \emph{non-quadratic}~\cite{Ciuciu99} and 2D (time-frequency) case would be of particular interest, and we are presently working at this issue. 

Whereas the first part of our contribution provides interpretations of pre-existing tools for spectral analysis, new estimation schemes are derived in the second part: unsupervised hyperparameter and window selection. It is shown that maximum likelihood solutions are both formally achievable and practically useful. 

%% file: annex.tex
\section{Proof of Propositions}
\label{AnnexProof}

\subsection{Proof of Proposition~\ref{PropUsualPenalCont}}
\label{DemPropUsualPenalCont}

Several proofs are available and the proposed one relies on variational principles~\cite{Luenberger69}. Application of these principles to quadratic regularization of linear problem yields the functional equation~\cite{Tikhonov77}:
\beqnx
-2 \WcFou^\dag (\yb-\WcFou a) + 2 \lambda I_{L^2} a = 0 \,,
\eeqnx
where $I_{L^2}$ stands for the identity application from $L^2$ onto itself and $\WcFou^\dag$ stands for the adjoint application of $\WcFou$ (see Appendix~\ref{AnnexAdjoint}). After elementary algebra we find:
\beqnx
(\WcFou^\dag \WcFou +\lambda I_{L^2} ) a = \WcFou^\dag \yb \,.
\eeqnx
As shown in Appendix~\ref{OrthoCont}, $\WcFou \WcFou^\dag = I_N$, then taking the FT and next the IFT gives:
\beqx
\est{a}^\lambda(\nu) = (1+\lambda)\pmu \WcFou^\dag \yb = (1+\lambda)\pmu \sum_{n=0}^{N-1} y_n e^{-2i\pi\nu n} \,.
\eeqx

\subsection{Proof of Proposition~\ref{PropUsualPenalDiscr}}
\label{DemPropUsualPenalDiscr}

The minimizer of the RLS criterion~(\ref{CritUsualPenalDiscr}) obviously is
\beqnx
\est{\ab}^\lambda &=& (\WFou^\dag \WFou +\lambda I_{\sss P} )\pmu \WFou^\dag \yb \,.
\eeqnx
One can refer to Appendix~\ref{AnnexDiagoDiscr} for a detailed calculus required to analyze the normal matrix $( \WFou^\dag \WFou + \lambda I_{\sss P} )$. $\WFou^\dag \WFou$ and $I_{\sss P}$ are circulant matrices, this property also holds for their sum which hence is diagonal in the Fourier basis. Elementary algebra leads to $\est{\ab}^\lambda$
\beqnx 
&=& \FFou \left[ \begin{array}{cc} 
		(1+\lambda)\pmu I_N	& O_{N,P-N} \\
		O_{P-N,N}				& \lambda\pmu I_{P-N}
				\end{array} \right] 
		\left[ \begin{array}{c} 
			I_N \\
			O_{P-N,N}
		\end{array} \right] \yb \\
&=&(1+ \lambda)\pmu \FFou \tilde{\yb}_P \,.
\eeqnx
%

\subsection{Proof of Proposition~\ref{PropWinPenalCont}}
\label{DemPropWinPenalCont}

The proof is founded on a time domain version of the criterion~(\ref{CritWinPenalCont}), resulting from application of the Plancherel-Parseval theorem to the successive derivatives of $a$:
\beqx 
\int_0^1 \left| \frac{\dd^q a}{\dd \nu^q}(\nu) \right|^2 \dd \nu = \sum_{n\in \eZ} (2\pi n)^{2q} |z_n|^2 \,,
\eeqx
where $z_n=\int_0^1 a(\nu) e^{2i\pi\nu n} \dd\nu$. Summation \wrt $q$ and inversion of summation \wrt $q$ and \wrt $n$, gives
\beqx
R_Q(a) = \sum_{n\in \eZ} e_n |z_n|^2 \,,
\eeqx
where the weighting coefficients $e_p$ fulfill~(\ref{DefVPWinPenalCont}). Hence, the time domain counterpart of criterion~(\ref{CritUsualPenalCont}) reads:
\beqx \label{CritUsualPenalContTemps}
\Qc_{\rm s}(a) = (\yb-\zb)^\dag (\yb-\zb) + \lambda \sum_{n\in\eZ} e_n |z_n|^2 \,.
\eeqx
Thanks to separability, the solution is easily derived: $\est{z}_n^\omega = (1+\lambda e_n)\pmu y_n$ if $n\in \eN_N$ and $\est{z}_n^\omega=0$ elsewhere. $a^\omega$ is the Fourier transform of the sequence $\{ \est{z}_n^\omega \}_{n\in \eZ}$
\beqx
\est{a}^\omega(\nu)= \sum_{n=0}^{N-1} \est{z}_n^\omega e^{-2i\pi\nu n} \,.
\eeqx

\subsection{Proof of Proposition~\ref{PropWinPenalDiscr}}
\label{DemPropWinPenalDiscr}

Elementary linear algebra provides the minimizer of~\ref{CritWinPenalDiscr}
\beqx
 \est{\ab}^\omega = (\WFou^\dag \WFou + \lambda \Pi_a )\pmu \WFou^\dag \yb \,.
\eeqx
Accounting for its circular structure, the Fourier basis diagonalizes $\Pi_a$:
\beqx
 \Pi_a = \FFou \Lambda_\Pi \FFou^\dag \,,
\eeqx
where $\Lambda_\Pi$ is the diagonal matrix of the eigenvalues $e_0,\dots,e_{P-1}$ of $\Pi_a$. Hence,
\beqx
\est{\ab}^\omega = \FFou (I_{\sss P} + \lambda \Lambda_\Pi) \tilde{\yb}_P \,,
\eeqx
and we easily find 
\beqx
\est{\ab}^\omega = \FFou \breve{\yb} \,,
\eeqx
with $\breve{y}_p = \omega_p \tilde{y}_p$ for $p\in \eN_P$, \ie the data vector windowed by 
\beqx
\omega_n = (1 + \lambda e_n)\pmu \,.
\eeqx

\subsection{Proof of Proposition~\ref{PropBayesCont}}
\label{DemPropBayesCont}

Let $\nu_0\in [0,1]$ and $a_0=a(\nu_0)$. Thanks to the linearity of the model~(\ref{ModelObserv}) and thanks to the Gaussian assumption for $a$ and $\bb$, the joint law of $(a_0,\yb)$ is also Gaussian. Hence, the random variable $(a_0\I\yb)$ is clearly Gaussian and it is well-known that its mean reads 
\beqx
\Esp{ a_0\I\yb } = R_{a_0\yb} R_{\yb}\pmu \yb \,,
\eeqx
where $R_{a_0\yb} =\Esp{a_0 \yb^\dag}$ and $R_{\yb} =\Esp{\yb \yb^\dag}$. Elementary algebra and independence of $a$ and $\bb$ yield 
\beqnx
R_{a_0\yb_n}	&=& \int_0^1 \Esp{a(\nu_0) a(\nu)^*} e^{-2i \pi \nu n} \dd\nu   + \Esp{a(\nu_0) b_n}	\\
					&=& \rond{\gamma_a}(n) e^{-2i \pi \nu_0 n} \,.
\eeqnx
%
%
Moreover, under the previously mentioned assumptions, the generic entry $R_{mn}$ for $R_{\yb}$ is  $R_{mn}=\Esp{y_m y_n^*}$
\beqn
	&=& \intdouble_0^1 \Esp{(a(\nu) a(\nu')^*}  e^{2i \pi (\nu n - \nu' m )} \dd\nu' \dd\nu + r_b \delta_{n-m} \nonumber\\
	&=& (\rond{\gamma_a}(n)+r_b) \,\delta_{n-m} \label{Rnm}\,,
\eeqn
%
%
where $\delta_n$ stands for the Kronecker sequence. Therefore, $R_{\yb}$ is a diagonal matrix with elements $(\rond{\gamma_a}(n)+r_b)$. Hence
\beqnx
\est{a}_0 &=& \sum_{n=0}^{N-1} \left[ 1+\lambda \rond{c_a}(n)\pmu \right]\pmu y_n e^{-2i\pi\nu_0 n}  \,,
\eeqnx
with $\lambda =r_b/r_a$. 
\section{Technical results}
\label{AnnexTech}

The present Appendix collects several useful properties of Fourier operators. In particular, special attention is paid to $\WFou$ and $\WcFou$. Some of the stated properties are classical. We have reported them in order to make our notations and normalization conventions explicit. The other properties are less usual, but all of them have straightforward proofs.

\subsection{Discrete case}

\subsubsection{Structure of $\FFou$} 

In the case of $N=P$, the matrix $\WFou$ identifies with the square matrix $\FFou^{\dag}$, where $\FFou$ performs the discrete FT for vectors of size $P$. We have the well-known orthogonality relations $\FFou^{\dag}\FFou=\FFou\FFou^{\dag}=I_P$ and $\FFou\T = \FFou$. 

\subsubsection{Structure of $W_{NP}$} 

The matrix $\WFou$ evaluates the FT on a discrete grid of $P$ points for sequences of $N$ points, $P \geqslant N$. Straightforward expansion of the product provides:
\beq \label{WFouFFou}
\WFou \FFou =	\left[ \begin{array}{cc} 
					I_N & O_{N,P-N} 
					\end{array} \right] \,.
\eeq
As a consequence, we obtain
\beq
\WFou^\dag \yb	= \FFou \left[ \begin{array}{c} 
							I_N \\ O_{P-N,N}				\label{Bourrage}
						\end{array} \right] \yb
					= \FFou \tilde{\yb}_P \,,
\eeq
where $\tilde{\yb}_P$ is the zero-padded version of \yb, up to length $P$.

\subsubsection{Structure of $\WFou^{\dag}\WFou$}
\label{AnnexDiagoDiscr}

The matrix $\WFou \WFou^{\dag}$ has a very simple structure since, for $P\geqslant N$: $\WFou\WFou^{\dag}=I_N$. Otherwise, $\WFou^{\dag} \WFou$ is a non-negative, Hermitian, $P\times P$ circulant matrix. Circularity results from digonalization in the Fourier basis $\FFou$: 
\beqnx
\WFou^{\dag} \WFou = \FFou \Lambda \FFou^{\dag} \,,
\eeqnx
and, from Eq.~(\ref{WFouFFou}):
\beqnx 
\Lambda
 &=& \left[ \begin{array}{cc} 
	 		I_N		& O_{N,P-N} 	\\
			O_{P-N,N}	& O_{P-N,P-N}
		\end{array} \right] \,.
\eeqnx
As a consequence, $\WFou^{\dag}\WFou$ has only two eigenvalues, $1$ and $0$, of respective order $N$ and $P-N$. Such a structure is useful in the proof of Propositions~(\ref{PropUsualPenalDiscr}) and~(\ref{PropWinPenalDiscr}) in Appendix~\ref{AnnexProof}.

%

\subsection{Continuous case}

\subsubsection{The $\WcFou$ operator} \label{AnnexAdjoint} \label{OrthoCont}

The linear application $\WcFou:~a\in L^2 \longrightarrow\zb\in \eC^N$ is defined  by $z_n=\int_0^1 a(\nu) e^{2i\pi\nu n} \dd\nu$ for $n\in \eN_N$. The adjoint operator $\WcFou^\dag:~\zb\in \eC^N\longrightarrow a=\WcFou^\dag\zb$ is the linear operator such that:
\beqx
\forall a\in L^2, \forall \zb \in \eC^N ~~~~
\langle \WcFou a , \zb \rangle_{\sss{\eC}^N} =  \langle a , \WcFou^\dag \zb \rangle_{\sss L^2} \,,
\eeqx
where $\langle\cdot,\cdot\rangle_{\sss{\eC}^N}$ and $\langle\cdot,\cdot\rangle_{\sss L^2}$ stand for the standard inner product in $\eC^N$ and $L^2$, respectively. It is given by:
\beqx
a(\nu) =\WcFou^\dag \zb = \sum_{n=0}^{N-1} z_n e^{-2i\pi\nu n} \,.
\eeqx
This can be justified as follows: by inverting the order of the finite sum $\sum_0^{N-1}$and the definite integral $\int_0^1$, we get
\beqx
\langle \WcFou a , \zb \rangle_{\sss{\eC}^N} = 
	\int_0^1 a(\nu) \sum_{n=0}^{N-1} z_n^* e^{2i\pi\nu n} = 
		\langle a , \WcFou^\dag \zb \rangle_{\sss L^2} \,.
\eeqx

Finally, elementary algebra shows that the composed application $\WcFou \WcFou^\dag$ is the identity application from $\eC^N$ onto itself.
%
%

\subsubsection{Technical results for the Example in~\ref{ExWinPenalContBayes}}

\paragraph{The Fourier series~(\ref{TheFourierSeries})}
\label{AnnexeFouSerie}

The proof of~(\ref{TheFourierSeries}) consists of three steps. The first one relies on the Fourier relationship between Cauchy and Laplace functions
\beqx
\frac{2\beta}{\beta^2+4\pi^2 t^2} = \int_{\eR} e^{-\beta |f|} ~e^{-2j\pi t f} \dd f \,, ~t\in\eR
\eeqx

The second step is founded on discrete time $t=n \in \eZ$ and expansion in a series of integrals:
\beqnx
\frac{2\beta}{\beta^2+4\pi^2 n^2}&=& \int_{\eR} e^{-\beta |f|} ~e^{-2j\pi n f} \dd f  \\
							  &=& \sum_{p\in\eZ} \int_0^1~e^{-\beta |\nu-p|}~e^{-2j\pi n\nu}\dd \nu   \\
							  &=& \int_0^1 \sum_{p\in\eZ} ~e^{-\beta |\nu-p|}~e^{-2j\pi n\nu} \dd \nu \,,\\
\eeqnx
%
%
since the invoked series are convergent. The last step is a simple geometric series calculus:
\beqx
\sum_{p\in\eZ} e^{-\beta |\nu-p|} = \frac{\cosh{\beta (\nu-1/2)}}{\sinh{\beta /2}} \,, \nu \in[0,1] \,,
\eeqx
easily obtained by rewriting the series as the sum of a series for $p\in\eZ_{-}$ (\ie $p\leqslant\nu$) and a series for $p\in\eZ_{+}^*$ (\ie $p\geqslant\nu$).

\paragraph{Conditional process}
\label{AnnexeCondProc}

Let us note $\nu,\nu'\in[0,1]$, $\nu>\nu'$. The partitioned vector $\bar{\ab}=[a(\nu),a(\nu'),a(1)]\T=[\tilde{\ab} | a_1]\T$ is clearly a zero-mean Gaussian vector with covariance
\beqx
R_{\bar{\ab}} = \left[\begin{matrix}
\gamma_a(0)				&\gamma_a(\nu-\nu')	&\gamma_a(\nu)		\\
\gamma_a(\nu-\nu')	&\gamma_a(0)			&\gamma_a(\nu')	\\
\gamma_a(\nu)			&\gamma_a(\nu')		&\gamma_a(0)
\end{matrix}\right] \,.
\eeqx
According to the conditional covariance matrix formula, $R_{\tilde{\ab}|a_1}=R_{\tilde{\ab}}-R_{\tilde{\ab} a_1}\T R_{a_1}\pmu R_{\tilde{\ab}a_1}$ we immediately get~(\ref{CondCorrel1}). Accounting for the explicit expression for $\gamma_a(\nu)$ given by~(\ref{Cov01}), simple expansion of hyperbolic functions yields~(\ref{CondCorrel2}).

\paragraph{Law of increments}
\label{AnnexeIncrements}

We have $\nu_1,\nu_2, \nu'_1,\nu'_2 \in[0,1]$, $\nu_1<\nu_2<\nu'_1<\nu'_2$. Let us introduce the collection of the four values $\underline{\ab}=[a(\nu_1),a(\nu_2), a(\nu'_1),a(\nu'_2)]$ which is clearly a zero-mean and Gaussian vector with covariance $R_{\underline{\ab}}$. The increment vector $\ib=[a(\nu_2)-a(\nu_1), a(\nu'_2)-a(\nu'_1)] \in \eC^2$ is a linear transform of the vector $\underline{\ab}$: $\ib=H\underline{\ab}$ with increment covariance $R_{\ib}$
\beqx
H = 
\left[\begin{matrix}
-1&1&0&0\\
0&0&-1&1
\end{matrix}\right] , \,
%
%
R_{\ib} = H R_{\underline{\ab}} H\T = 
\left[\begin{matrix}
r_i	&\rho	\\
\rho	&r'_i	\\
\end{matrix}\right] \,.
\eeqx
with $r_i = 2(\gamma_a(0)-\gamma_a(\nu_2-\nu_1))$, $r'_i = 2(\gamma_a(0)-\gamma_a(\nu'_2-\nu'_1))$, and $\rho = \gamma_a(\nu_2-\nu'_2)+\gamma_a(\nu_1-\nu'_1)  -\gamma_a(\nu_1-\nu'_2)-\gamma_a(\nu_2-\nu'_1)$. Finally, Taylor development at $\alpha_0=0$ yields $r_i  = (\nu_2-\nu_1) (1-(\nu_2-\nu_1))/2\alpha_1$, $r'_i = (\nu'_2-\nu'_1) (1-(\nu'_2-\nu'_1))/2\alpha_1$, and $\rho = (\nu_2-\nu_1) (\nu'_2-\nu'_1)/\alpha_1$, and proves~(\ref{IncreCov}).

\section*{Acknowledgement}

First author is particularly thankful to Alain, Naomi, Philippe and Denise for committed support and coaching.

%% file: main.bbl
\begin{thebibliography}{10}

\bibitem{Robinson82}
E.~R. Robinson,
\newblock ``A historical perspective of spectrum estimation'',
\newblock {\em {P}roc. \uppercase{ieee}}, vol. 9, no. 9, pp. 885--907,
  {S}eptember 1982.

\bibitem{Kay81}
S.~M. Kay and S.~L. Marple,
\newblock ``Spectrum analysis~--~a modern perpective'',
\newblock {\em {P}roc. \uppercase{ieee}}, vol. 69, no. 11, pp. 1380--1419,
  {N}ovember 1981.

\bibitem{Marple87}
S.~L. Marple,
\newblock {\em Digital Spectral Analysis with Applications},
\newblock Prentice-Hall, Englewood Cliffs, \sca{nj}, 1987.

\bibitem{Kay88}
S.~M. Kay,
\newblock {\em Modern Spectral Estimation},
\newblock Prentice-Hall, Englewood Cliffs, \sca{nj}, 1988.

\bibitem{Sacchi98}
M.~D. Sacchi, T.~J. Ulrych, and C.~J. Walker,
\newblock ``Interpolation and extrapolation using a high-resolution discrete
  {F}ourier transform'',
\newblock {\em \uppercase{ieee} {T}rans. {S}ignal {P}rocessing}, vol. 46, no.
  1, pp. 31--38, {J}anuary 1998.

\bibitem{Sacchi96}
M.~D. Sacchi and T.~J. Ulrych,
\newblock ``Estimation of the discrete {F}ourier transform, a linear inversion
  approach'',
\newblock {\em Geophysics}, vol. 61, no. 4, pp. 1128--1136, 1996.

\bibitem{Sorenson80}
H.~W. Sorenson,
\newblock {\em Parameter estimation}, vol.~9 of {\em Control and system
  theory},
\newblock Marcel Dekker, New York Basel, 1980.

\bibitem{Tikhonov77}
A.~Tikhonov and V.~Arsenin,
\newblock {\em Solutions of Ill-Posed Problems},
\newblock Winston, Washington, \sca{dc}, 1977.

\bibitem{Nashed74}
M.~Z. Nashed and G.~Wahba,
\newblock ``Generalized inverses in reproducing kernel spaces: An approach to
  regularization of linear operators equations'',
\newblock {\em \uppercase{siam} {J}. {M}ath. {A}nal.}, vol. 5, pp. 974--987,
  1974.

\bibitem{Demoment89}
G.~Demoment,
\newblock ``Image reconstruction and restoration: Overview of common estimation
  structure and problems'',
\newblock {\em \uppercase{ieee} {T}rans. {A}coust. {S}peech, {S}ignal
  {P}rocessing}, vol. \sca{assp}-37, no. 12, pp. 2024--2036, {D}ecember 1989.

\bibitem{Kitagawa85}
G.~Kitagawa and W.~Gersch,
\newblock ``A smoothness priors long {AR} model method for spectral
  estimation'',
\newblock {\em \uppercase{ieee} {T}rans. {A}utomat. {C}ontr.}, vol.
  \sca{ac}-30, no. 1, pp. 57--65, {J}anuary 1985.

\bibitem{Wahba80}
G.~Wahba,
\newblock ``Automatic smoothing of the log periodogram'',
\newblock {\em J. of the American Statistical Association, Theory and Methods
  Section}, vol. 75, no. 369, pp. 122--132, {M}arch 1980.

\bibitem{Bretthorst88}
G.~L. Bretthorst,
\newblock {\em {B}ayesian Spectrum Analysis and Parameter Estimation}, vol.~48
  of {\em Lecture Notes in Statistics},
\newblock J. Berger, S.Fienberg, J. Gani, K. Krickeberg, and B. Singer,
  {S}pringer-{V}erlag edition, 1988.

\bibitem{Dublanchet97}
F.~Dublanchet, J.~Idier, and P.~Duvaut,
\newblock ``Direction-of-arrival and frequency estimation using
  {P}oisson-{G}aussian modeling'',
\newblock in {\em {P}roc. \uppercase{ieee} \uppercase{icassp}}, Munich,
  Germany, {A}pril 1997, pp. 3501--3504.

\bibitem{Harris78}
F.~J. Harris,
\newblock ``On the use of windows for harmonic analysis with the discrete
  {F}ourier transform'',
\newblock {\em {P}roc. \uppercase{ieee}}, vol. 66, no. 1, pp. 51--83, {J}anuary
  1978.

\bibitem{Bertin93}
A.~Bertin,
\newblock {\em Espaces de Hilbert},
\newblock Service d'\'Edition de l'ENSTA, 1993.

\bibitem{Cramer67}
H.~Cram\'er and M.~R. Leadbetter,
\newblock {\em Stationary and {R}elated {S}tochastic {P}rocesses},
\newblock John Wiley, New York, London, Sydney, 1967.

\bibitem{Bremaud99}
P.~Br{\'e}maud,
\newblock {\em {M}arkov Chains. {G}ibbs fields, Monte Carlo Simulation, and
  Queues},
\newblock Texts in Applied Mathematics~31. Spinger, New York, \sca{ny}, 1999.

\bibitem{Moura97}
J.~M.~F. Moura and G.~Sauraj,
\newblock ``{G}auss-{M}arkov random fields ({GM}rf) with continuous indices'',
\newblock {\em \uppercase{ieee} {T}rans. {I}nf. {T}heory}, vol. 43, no. 5, pp.
  1560--1573, {S}eptember 1997.

\bibitem{Wong71}
E.~Wong,
\newblock {\em Stochastic Processes in Information and Dynamical Systems},
\newblock Series in Systems Science. McGraw-Hill Book Company, New York,
  \sca{ny}, 1971.

\bibitem{Bhattacharya90}
R.~N. Bhattacharya and E.~C. Waymire,
\newblock {\em Stochastic Processes with Applications},
\newblock John Willay \& Sons, Inc., New York, \sca{ny}, 1990.

\bibitem{Golub79}
G.~H. Golub, M.~Heath, and G.~Wahba,
\newblock ``Generalized cross-validation as a method for choosing a good ridge
  parameter'',
\newblock {\em {T}echnometrics}, vol. 21, no. 2, pp. 215--223, {M}ay 1979.

\bibitem{Titterington85}
D.~M. Titterington,
\newblock ``Common structure of smoothing techniques in statistics'',
\newblock {\em {I}nt. {S}tatist. {R}ev.}, vol. 53, no. 2, pp. 141--170, 1985.

\bibitem{Hall87}
P.~Hall and D.~M. Titterington,
\newblock ``Common structure of techniques for choosing smoothing parameter in
  regression problems'',
\newblock {\em {J}. {R}. {S}tatist. {S}oc. B}, vol. 49, no. 2, pp. 184--198,
  1987.

\bibitem{Thompson91}
A.~Thompson, J.~C. Brown, J.~W. Kay, and D.~M. Titterington,
\newblock ``A study of methods of choosing the smoothing parameter in image
  restoration by regularization'',
\newblock {\em \uppercase{ieee} {T}rans. {P}attern {A}nal. {M}ach. {I}ntell.},
  vol. \sca{pami}-13, no. 4, pp. 326--339, {A}pril 1991.

\bibitem{Fortier92}
N.~Fortier, G.~Demoment, and Y.~Goussard,
\newblock ``Comparison of \sca{gcv} and \sca{ml} methods of determining
  parameters in image restoration by regularisation'',
\newblock {\em {J}. {V}isual {C}omm. {I}mage {R}epres.}, vol. 4, pp. 157--170,
  1993.

\bibitem{Giovannelli96}
J.-F. Giovannelli, G.~Demoment, and A.~Herment,
\newblock ``A {B}ayesian method for long \sca{ar} spectral estimation: {a}
  comparative study'',
\newblock {\em \uppercase{ieee} {T}rans. {U}ltrasonics {F}erroelectrics and
  {F}requency {C}ontrol}, vol. 43, no. 2, pp. 220--233, {M}arch 1996.

\bibitem{Bertsekas95}
D.~P. Bertsekas,
\newblock {\em Nonlinear programming},
\newblock Athena Scientific, Belmont, \sca{ma}, 1995.

\bibitem{Shumway82}
R.~Shumway and D.~Stoffer,
\newblock ``An approach to time series smoothing and forecasting using the
  \sca{em} algorithm'',
\newblock {\em {J}. {T}ime {S}eries {A}nalysis}, pp. 253--264, 1982.

\bibitem{Basseville89}
M.~Basseville,
\newblock ``Distance measures for signal processing and pattern recognition'',
\newblock {\em {S}ignal {P}rocessing}, vol. 18, no. 4, pp. 349--369, {D}ecember
  1989.

\bibitem{Ciuciu99}
P.~Ciuciu, J.~Idier, and J.-F. Giovannelli,
\newblock ``{M}arkovian high resolution spectral analysis'',
\newblock in {\em {P}roc. \uppercase{ieee} \uppercase{icassp}}, Phoenix,
  \sca{az}, {M}arch 1999, pp. 1601--1604.

\bibitem{Luenberger69}
D.~G. Luenberger,
\newblock {\em Optimization by Vector Space Methods},
\newblock John Wiley, New York, \sca{ny}, 1st edition, 1969.

\end{thebibliography}
